\documentclass[aps,prl,twocolumn,nofootinbib,notitlepage,superscriptaddress,aps,pra]{revtex4-1}
\usepackage{float}
\usepackage{amsmath}
\usepackage{amssymb}
\usepackage{graphicx}
\usepackage{soul}
\usepackage{physics}
\usepackage{color}
\usepackage{mathtools}
\usepackage[colorlinks=true,linkcolor=black,anchorcolor=black,citecolor=black,filecolor=black,menucolor=black,runcolor=black,urlcolor=black]{hyperref}
\usepackage{cleveref}
\usepackage{tabularx}

\begin{document}

\title{Transverse relative locality effects in de Sitter spacetime}

\author{Giuseppe Fabiano}
\affiliation{Physics Division, Lawrence Berkeley National Laboratory, Berkeley, CA 94720, USA}
\affiliation{Department of Physics, University of California, Berkeley, CA 94720, USA}
\affiliation{Centro Ricerche Enrico Fermi, I-00184 Rome, Italy}
\author{Domenico Frattulillo}
\affiliation{Istituto Nazionale di Fisica Nucleare, Sezione di Napoli, Complesso Univ. Monte S. Angelo, I-80126 Napoli, Italy}

\begin{abstract}
Doubly Special Relativity (DSR) models are characterized by the deformation of relativistic symmetries at the Planck scale and constitute one of the cornerstones for quantum gravity phenomenology research, due to the possibility of testing them with cosmological messengers. Some of their predictions manifest themselves as relative locality effects, implying that events local to an observer might not appear to be so for a distant one. In this work we focus on transverse relative locality models, where the delocalization occurs along the direction perpendicular to the one connecting two distant observers. We present the first generalization of these models in curved spacetime, constructing a transverse deformation of the de Sitter algebra in $2+1$ D and investigating its phenomenological implications on particle propagation.
\end{abstract}

\maketitle

Studies on possible departures from local Lorentz invariance at the Planck scale represent one of the most active areas of research in quantum gravity phenomenology \cite{Amelino-Camelia:2008aez,Mattingly:2005re,Addazi:2021xuf}. This interest is driven by the possibility to test Planck scale effects through the analysis of particle propagation over cosmological distances, which can act as huge amplifiers, making even tiny deviations from standard relativistic predictions potentially observable \cite{Amelino-Camelia:1997ieq}. The phenomenological models proposed in the literature either break Lorentz symmetry (LIV) \cite{Amelino-Camelia:1997ieq,Alfaro:1999wd,Gambini:1998it} or preserve the equivalence of inertial reference frames at the cost of deforming relativistic symmetries (DSR) \cite{Amelino-Camelia:2000stu,Amelino-Camelia:2002cqb,Magueijo:2002am,Kowalski-Glikman:2002iba}. The latter case introduces an invariant energy scale, assumed to be of the order of the Planck energy, on par with the speed of light. In this letter we focus on a much studied aspect of DSR models known as relative locality, implying that events which appear to be local to nearby observers might appear not to be so for distant ones \cite{Amelino-Camelia:2011lvm,Amelino-Camelia:2011hjg,Arzano:2010kz,Smolin:2010mx}. This departure from absolute locality is analogous to the loss of absolute simultaneity when transitioning from Galilean to special relativity, in order to accommodate the speed of light as an invariant velocity scale. The majority of studies in this field have focused on longitudinal relative locality, where the delocalization effects, characterized in both flat and curved spacetime, occur along the direction connecting the observers \cite{Amelino-Camelia:2010wxx,Amelino-Camelia:2012vzf,Rosati:2015pga,Amelino-Camelia:2023srg}.

In recent years there has also been a growing interest in transverse delocalization effects \cite{Freidel:2011mt,Amelino-Camelia:2011ycm,Amelino-Camelia:2017pne,Steinbring:2006ja,Steinbring:2023abf,Lee:2023kry} (which instead occur along the direction orthogonal to the one connecting the observers), especially in light of the development of the GrailQuest and HERMES space missions \cite{GrailQuest:2019ood,HERMES-SP:2021mwp,Ghirlanda:2024ayb}, whose ambition is to deploy a constellation of micro-satellites capable of identifying astrophysical sources with very high angular resolution. The predictions of Refs. \cite{Freidel:2011mt,Amelino-Camelia:2011ycm,Amelino-Camelia:2017pne} fall within the DSR framework and are purely systematic effects manifesting themselves as energy-dependent spatial shifts and angular deviations of the trajectories of high-energy particles as inferred by an observer distant from the source that is emitting them. On the other hand, the predictions of Refs. \cite{Steinbring:2006ja,Steinbring:2023abf,Lee:2023kry} are spacetime fuzziness effects which manifest themselves in the blurring of the image of distant astrophysical sources.

In this work we will focus on the former class of effects, which up to now have only been studied in the flat space-time scenario. Of course, this is a key limitation for phenomenological analyses given that spacetime curvature and expansion are clearly relevant over cosmological distances. We take the first step towards addressing this issue by presenting an investigation of transverse relative locality effects in de Sitter spacetime. For concreteness, we focus on the $2+1$ D case, but we expect the generalization to the $3+1$ D case to be straightforward. In the following, we work in units in which $\hbar=c=1$.

Our starting point is an ansatz for a “transverse” modification of the de Sitter symmetry algebra in $2+1$ D. In the following, $H$ denotes the de Sitter spacetime curvature parameter and $\ell$ an inverse energy scale of the order of the inverse of the Planck energy. For phenomenological purposes we will work at first order in $\ell$. Our ansatz reads 
\begin{equation}
\label{eq:algebradef}
    \begin{aligned}
        \{E,P_i\}=&H P_i +a \ell H \epsilon_{ij}EP_j+a \ell H^2 R P_i \, ,  \\ \{P_i,P_j\}=&-b\ell H  \epsilon_{ij}\vec{P}^2 \, ,\\ 
        \{N_i,P_j\}=&\delta_{ij}E+H \epsilon_{ij}R-a \ell H \delta_{ij}R E+\\
        -&b \ell \epsilon_{jk}P_iP_k+b\ell \epsilon_{ij}E^2+\\ 
        -&b \ell H \epsilon_{jk} N_i P_k-b \ell H \delta_{ij}\epsilon_{hk}N_hP_k\,,\\
        \{N_1,N_2\}=&-R\, ,\\
        \{N_i,E\}=&P_i+H N_i+b\ell \epsilon_{ij}EP_j-a\ell H \epsilon_{ij} E N_j+\\+&a\ell H R P_i+a\ell H^2 R N_i\,,\\ \{R,P_i\}=&\epsilon_{ij}P_j\, ,\\
        \{R,N_i\}=&\epsilon_{ij}N_j\,,\\\{R,E\}=&0 \,,
    \end{aligned}
\end{equation}
where $a,b$ are dimensionless parameters. 
The deformations in \eqref{eq:algebradef} are constructed such that the generators satisfy the Jacobi identities. Also, in the limit $\ell\rightarrow 0$ one recovers the de Sitter algebra in 2+1 D and in the limit $\ell\rightarrow0,\,H\rightarrow0$ one recovers the 2+1 D Poincaré algebra.

Finally, it can be shown that the deformed algebra in \eqref{eq:algebradef} leaves invariant the following Casimir element:
\begin{equation}
\begin{aligned}
\label{eq:casimir}
    C=&E^2-\vec{P}^2-2H \vec{N}\cdot \vec{P}-H^2 R^2+\\-&2(a+b)\ell H \epsilon_{ij}E N_iP_j \, .
    \end{aligned}
\end{equation}
We now introduce spacetime coordinates and their conjugate variables in order to describe particle worldlines. Similar to past works \cite{Rosati:2015pga,Amelino-Camelia:2012vzf} dealing with deformations of kinematics in de Sitter spacetime, we find it convenient to adopt conformal coordinates, in which the conformal time $\eta$ is related to the comoving time $t$ by 
\begin{equation}
    \eta=\frac{1-e^{-Ht}}{H} \, .
\end{equation}
The spacetime coordinates $(\eta,x_i)$ and their canonically conjugate variables $(\Omega,\Pi_i)$ satisfy 
\begin{equation}
\label{eq:conjvar}
    \begin{aligned}
        &\{\Omega,\eta\}=1,\;\;\; \{\Omega,x_i\}=0,\;\;\;\{\Omega,\Pi_i\}=0,\\
        &\{x_i,\eta\}=0,\;\;\; \{\Pi_i,x_j\}=-\delta_{ij},\;\;\;\{\Pi_i,\eta\}=0 \, .
    \end{aligned}
\end{equation}
In terms of these phase space variables, we provide a representation of the symmetry generators introduced in \eqref{eq:algebradef}:
\begin{equation}
\label{eq:genrepr}
    \begin{aligned}
        E=& \Omega+H (-\eta  \Omega +\vec x\cdot\vec \Pi )+\\+&a\ell  H\epsilon_{ij}x_i\Pi_j(H (-\eta  \Omega +\vec x\cdot\vec \Pi)+\Omega),\\
        P_i=&\Pi_i-b\ell   \epsilon_{ij}\Pi_j(\Omega+H (-\eta  \Omega +\vec x\cdot\vec \Pi))\, ,\\
        N_i=&x_i( \Omega+H (-\eta  \Omega +\vec x\cdot\vec \Pi ))+\\-&\Pi_i(\eta -\frac{1}{2} H \left(\eta ^2-\vec{x}^2\right))\, ,\\
        R=&\epsilon_{ij}x_i\Pi_j \, ,
    \end{aligned}
\end{equation}
as well as of the Casimir element \eqref{eq:casimir}:
\begin{equation}
\label{eq:casimirrep}
    C=(1-H\eta  )^2 \left(\Omega ^2-\vec{\Pi}^2\right) \, .
\end{equation}
We work within a covariant Hamiltonian formulation of classical mechanics
\cite{Rosati:2015pga,Amelino-Camelia:2012vzf} in which the Hamiltonian constraint is given by $\mathcal{H}=C-m^2=0$. Our phenomenological analysis will focus on massless particles, whose worldlines are found by simply setting $m=0$ in the constraint: 
\begin{equation}
\label{eq:pworldline}
    \vec{x}-\vec{x}_O=\frac{\vec{\Pi}}{|\vec{\Pi}|}(\eta-\eta_O) \, ,
\end{equation}
where $\vec x_O$ is the position of the particle when $\eta=\eta_O$.

Consider now the physical scenario in which an observer, Alice, is local to a source that simultaneously emits two massless particles traveling along the $x_1$ direction. We will assume that one of the particles is “soft" and the other is “hard", in the sense that Planck scale deformation effects on the former are negligible with respect to the latter. In Alice's reference frame, the worldlines can be written as
\begin{equation}
\label{eq:Alicewl}
\begin{aligned}
    &x_{1,h}^A=\eta^A_h \, , \\
&x_{2,h}^A=0 \, , \\
    &x_{1,s}^A=\eta^A_s \, , \\
&x_{2,s}^A=0 \, ,
\end{aligned} 
\end{equation}
where the subscripts “$h$" and “$s$" refer to the hard and soft particle, respectively, the superscript “$A$" refers to Alice, and we have assumed $\Pi_1>0$. We want to map Alice's description of the worldlines \eqref{eq:Alicewl} to the description of an observer local to a detector, Bob, connected to Alice by a finite spatial translation along $x_1$, followed by a finite time translation. 
The relevant phase space variables in Bob's reference frame are related to Alice's one by means of the relation
\begin{equation}
\label{eq:BobAlicerel}
(\eta,x_i,\Omega,\Pi_i)^B=e^{-{\xi_1}{P_1}}\rhd e^{-\zeta E}\rhd(\eta,x_i,\Omega,\Pi_i)^A \, ,
\end{equation}
where $\rhd$ represents the action by Poisson bracket of the corresponding generators\footnote{For a transformation identified by a  generator $G$ and a parameter $a$, the finite action on a coordinate $X$ is $e^{{a}G} \triangleright X \equiv \sum_{n=0}^{\infty} \frac{a^n}{n!}\left\lbrace G,X\right\rbrace_n $, where $ \lbrace G,X \rbrace_n = \lbrace G, \lbrace G,X  \rbrace_{n-1}\rbrace$, $\lbrace G, X \rbrace_0 = X $.
In this formalism, a spatial translation along $x_1$ followed by a time translation is given by $e^{-\xi_1 P_1}\triangleright e^{-{\zeta}E}\triangleright X$.} and $\xi_1$ and $\zeta$
are respectively the space and time translation parameters. In particular, in order for the worldline of the soft photon to cross Bob's spacetime origin, we choose the translation parameters in the following way
\begin{equation}
    \zeta=T,\qquad \xi_1=\frac{1-e^{-HT}}{H} \, ,
\end{equation}
where $T$ is the comoving distance between Alice and Bob. Using relations \eqref{eq:conjvar},\eqref{eq:genrepr} and \eqref{eq:BobAlicerel}, the worldlines in Bob's reference frame are thus written as

\begin{equation}
\label{eq:Bobwl}
\begin{aligned}
    &x_{1,h}^B=\eta^B_h \, ,\\
&x_{2,h}^B=b\frac{\ell }{H}\Pi^B_{1,h}e^{H T}\sinh{(H T)}+a \ell H T   \Pi^B_{1,h} \eta^B_h\, ,\\
    &x_{1,s}^B=\eta^B_s \, ,\\
    &x_{2,s}^B=0 
 \, .
\end{aligned}
\end{equation}

\begin{figure*}[t!]
    \centering
    \includegraphics[scale=1.05]{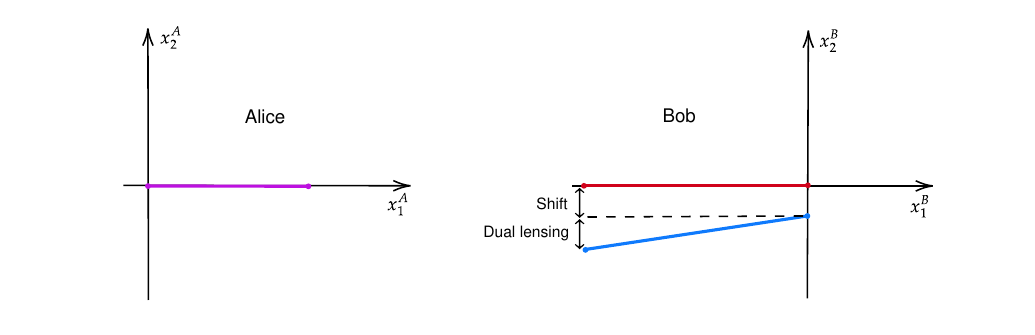}
    \caption{Pictorial representation of the geometry of particle worldlines according to observers Alice and Bob, connected by a time translation and a space translation along the $x_1$ direction. On the left-hand side we have Alice's reference frame in which the worldlines of the soft and hard particle overlap and lie entirely on the $x_1$-axis. The worldlines are depicted in purple and their end-points identify the emission (left) and detection (right) according to Alice. In Bob's reference frame, transverse relative locality effects manifest themselves in the form of a shift and an angular deviation of the worldlines of the hard (blue) and soft (red) particle. According to Bob, both the emission and the detection of the particles is non-local, pictorially characterized by non-coincident end-points identifying the emission (left) and detection (right).}
    \label{fig}
\end{figure*}

As required, the worldline of the soft particle crosses Bob's spacetime origin. However, the transverse symmetry deformations allow for a non-vanishing expression for $x_{2,h}^B$, so that the hard particle misses Bob's spatial origin when $\eta^B=0$. Quantitatively, this shift is expressed by
\begin{equation}
\label{eq:shift}
\begin{aligned}
    \Delta x_2^B=&b\frac{\ell}{H}\Pi^B_{1,h}e^{H T}\sinh{ H T}=\\
    =&\frac{b \ell \Pi_{1
    ,h}^B z (z+2)}{2 H} \, ,
    \end{aligned}
\end{equation}
which we have also parametrized in terms of the redshift of the source $z=e^{HT}-1$. Notice that Bob could detect both particles only if he were equipped with another detector, appropriately displaced with the one local to his origin. Inspecting the hard particle worldline expression in \eqref{eq:Bobwl}, we also see that the worldline forms a non-zero angle with respect to the Bob's $x_1$ axis, even though the particles were emitted parallel to each other and along the $x_1$ axis according to Alice. This angular deviation is refereed to as a dual lensing effect \cite{Freidel:2011mt,Amelino-Camelia:2011ycm} and is given by
\begin{equation}
\label{eq:lensing}
    \Delta\theta=a\ell H\Pi_{1,h}^B T=a\ell  \Pi_{1,h}^B\log{(1+z)} \, ,
\end{equation}
and is amplified by the distance between the observers. It is also interesting to observe that the dual lensing in \eqref{eq:lensing} is a curvature-induced effect, in the sense that it vanishes in the limit $H\rightarrow 0$ \cite{Amelino-Camelia:2020bvx,Amelino-Camelia:2023srg}. In the case of purely translated observers, the curvature-induced nature of the dual lensing effect is to be expected from a simple dimensional argument. By construction, relative locality effects will be proportional to the product of the deformation scale $\ell$ and the momentum of the particle involved. An angular deviation effect that takes the relative distance between two observers into account must also depend on the translation parameter, which can only form a dimensionless quantity when multiplied by an inverse length scale, such as the curvature scale $H$. This is in contrast with the flat spacetime version of the effect, derived in \cite{Freidel:2011mt,Amelino-Camelia:2011ycm}, where the absence of $H$ makes it so that it is impossible to obtain a Planck scale angular delocalization in the case of purely translated observers, and indeed the effect of Refs. \cite{Freidel:2011mt,Amelino-Camelia:2011ycm} is present only when the observers are also relatively boosted. In that case there is no dependence on the translation parameter that connects the observers, so the effect is not amplified by the distance. On the other hand, the shift effect in \eqref{eq:shift} is also present in the limit $H\rightarrow 0$ and is amplified by the distance between observers also in the flat spacetime limit.

The implications of the dual lensing and shift effects on the shape of particle worldlines as described by Bob are depicted in \Cref{fig}, which shows how relative locality manifests itself as a misleading inference by Bob on the source of the particles. Indeed, while according to Alice the two particles are emitted from the same spacetime point, according to \Cref{fig}, Bob will deduce that the particles are actually emitted from different spacetime points.

The predictions of Eqs. \eqref{eq:shift} and \eqref{eq:lensing} are the main results of our study and show for the first time the interplay between Planck scale and spacetime curvature in transverse relative locality effects. To refine this model, a future goal is to generalize it to 3+1 dimensions and to take into account the non-constant rate of expansion of the universe, characterizing transverse relative locality effects in FLRW spacetime. This will allow us to establish a stronger connection between our theoretical predictions and cosmological observations, particularly those originating from sources at very high redshift distances for which the constant rate of expansion approximation becomes less reliable. Nonetheless, we believe that with this study we have filled an important gap in the transverse relative locality literature accounting for the role of spacetime curvature and paving the way for a quite unexplored phenomenological area.

\section*{Acknowledgements}
We thank Giovanni Amelino-Camelia for useful comments which helped to improve the quality of this manuscript.
We would also like to acknowledge the contribution of the COST Action CA23130 ``Bridging high and low energies in search of quantum gravity (BridgeQG)''.
G.F. acknowledges financial support from the ``Foundation Blanceflor''.

\bibliographystyle{bibliostyle}
\bibliography{Refs}

\end{document}